\documentclass[11pt,a4paper]{article}
\usepackage{jheppub}
\usepackage{amsmath}
\usepackage{mathtools}
\usepackage{amssymb,bm,bbold}
\usepackage{enumerate}
\usepackage{comment}
\usepackage{esvect}
\usepackage{import}
\usepackage{subcaption}
\usepackage{shuffle}
\usepackage{flexisym}
\usepackage{breqn}
\usepackage{fancyhdr}
\usepackage{tikz}

  \usetikzlibrary {positioning}
  \usetikzlibrary{shapes.geometric, arrows}
	 \usetikzlibrary{arrows,decorations.markings}
	 \usetikzlibrary{calc}
	 \usetikzlibrary{3d}

\usepackage{tikz-feynman}
\tikzfeynmanset{compat=1.1.0}
 
\usepackage{feynmp}
\DeclareGraphicsExtensions{.mps}
\usepackage{graphicx}
\usepackage{pgfplots}
\pgfplotsset{compat=newest}
\usepgfplotslibrary{fillbetween}
\usepackage{cancel}

\newtheorem{definition}{Definition}

\newcommand{\dlog}{d\text{log}}
\newcommand{\be}{\begin{equation}}
\newcommand{\ee}{\end{equation}}

\newcommand{\dd}{\text{d}}

\def\cN{{\cal N}}

\newcommand{\rigers}[1]{{\color{orange}[\textbf{Rigers:} #1]}}
\title{An Exceptional Cluster Algebra for Higgs plus Jet Production}
\author[a,b]{Rigers Aliaj}
\author[b,c]{and Georgios Papathanasiou}
\affiliation[a]{ II. Institut f{\"u}r Theoretische Physik, Universit{\"a}t Hamburg, Luruper Chaussee 149, 22607 Hamburg,
Germany}
\affiliation[b]{Deutsches Elektronen-Synchrotron DESY, Notkestr. 85, 22607 Hamburg, Germany}
\affiliation[c]{Department of Mathematics, City, University of London,
Northampton Square, EC1V 0HB, London, UK}

\emailAdd{rigers.aliaj@desy.de}
\emailAdd{georgios.papathanasiou@desy.de}
\abstract{A recent evaluation of three-loop nonplanar Feynman integrals contributing to Higgs plus jet production has established their dependence on two novel symbol letters. We show that the resulting alphabet is described by a $G_2$ cluster algebra, enlarging the $C_2$ cluster algebra found to cover all previously known integrals relevant for this process. The cluster algebra connection we find reveals new adjacency relations, which significantly reduce the function space dimension of the non-planar triple ladder integral. These adjacencies may be understood in part by embedding $G_2$ inside higher-rank cluster algebras.}
\arxivnumber{2408.14544 }
\begin{document}
\maketitle
\flushbottom

\section{Introduction}

Following the discovery of the Higgs boson, a new era of precision measurements has begun at the Large Hadron Collider and its planned High-Luminosity upgrade. Interpreting these measured observables, determining the parameters of the Standard Model as well as telling apart the subtle signature of new physics from them, requires their theoretical description to reach a commensurate level of accuracy. One of the key challenges to this end, remains the computation of scattering amplitudes and their building blocks, Feynman integrals, at higher orders in perturbative quantum field theory~\cite{Huss:2022ful}.

On the analytic front, that often provides a faster and stabler evaluation than the numeric one, the method of choice for computing the  master integrals (namely the solutions of the linear integration-by-parts identities~\cite{Chetyrkin:1981qh} among all Feynman integrals contributing to a given process) are differential equations~\cite{Kotikov:1990kg,Remiddi:1997ny,Gehrmann:1999as} in canonical form~\cite{Henn:2013pwa}. Working in dimensional regularisation, where the dimension of loop momenta is $D=4-2\epsilon$, and focusing on bases of integrals $\mathbf{f}$ which evaluate to the often sufficient class of \emph{multiple polylogarithms} \cite{Chen1977,goncharov2001multiple,Goncharov2005}, these canonical differential equations take the form
\begin{equation}\label{eq:CDE_intro}
    d\mathbf{f}(\vec{z};\epsilon)=\epsilon \left[\sum_{i}\mathbf{A}_{i} \,\dlog\alpha_i(\vec{z})\right]\mathbf{f}(\vec{z};\epsilon).
\end{equation}
Here, $\vec{z}$ collectively denotes the kinematic variables the integrals depend on, such as external momenta and internal masses, and $d=\sum_{j} \dd z_{j} \partial_j$ is the total differential. Finally, each $\alpha_i$ is an algebraic function of the $\vec{z}$ components known as a \emph{letter} (of the symbol~\cite{Goncharov:2010jf}), with the entire set $\mathcal{A}\equiv\{\alpha_i\}$ similarly denoted as the (symbol) \emph{alphabet}, and $\mathbf{A}_i$ are constant matrices. 

Despite the great success of the method, for a state-of-the-art application see e.g.~\cite{Abreu:2023rco}, this too becomes increasingly unwieldy as the perturbative order and number of kinematic variables grow to meet experimental demands.  Serious bottlenecks include analytically solving the integration-by-parts identities in terms of an initial basis, as well as determining the basis transformation that brings it to the form~\eqref{eq:CDE_intro}. However prior knowledge of the alphabet can be extremely helpful with these calculations, as it converts them from symbolic to much simpler, numeric ones~\cite{Abreu:2018rcw}. This fact renders the prediction of the alphabet by independent means an attractive endeavour.

In this respect, mathematical objects known as \emph{cluster algebras}~\cite{1021.16017} appear quite promising. Cluster algebras have been first observed to describe the alphabet of six- and seven-particle amplitudes in planar $\cN=4$ super Yang-Mills theory (SYM)~\cite{Golden:2013xva}, providing crucial information for computing these amplitudes to unprecedented loop orders by bootstrap methods~\cite{Dixon:2011pw,Dixon:2011nj,Dixon:2013eka,Dixon:2014voa,Dixon:2014iba,Drummond:2014ffa,Dixon:2015iva,Caron-Huot:2016owq,Dixon:2016nkn,Drummond:2018caf,Caron-Huot:2019vjl,Caron-Huot:2019bsq,Dixon:2020cnr,Chestnov:2020ifg,Dixon:2023kop}\footnote{Before that, cluster algebras also appeared at the level of the amplitude integrand in this theory~\cite{Arkani-Hamed:2012zlh}, as well as enjoyed remarkable connections to other theoretical physics topics, for example thermodynamic Bethe ans\"atze~\cite{Fomin:2001rc}, moduli spaces~\cite{Gaiotto:2009hg}  and electric/magnetic duality~\cite{Feng:2001bn} of supersymmetric gauge theories, or mirror symmetry~\cite{2013arXiv1309.2573G}.}; and closely related generalisations are also seen to describe higher-point amplitudes of the same theory~\cite{Drummond:2019cxm,Arkani-Hamed:2019rds,Henke:2019hve,Herderschee:2021dez,Henke:2021ity,Ren:2021ztg}. Most importantly, in~\cite{Chicherin:2020umh} it was discovered that cluster-algebraic structures are not confined to idealised models: In particular, it was demonstrated that cluster algebras also underlie the analytic structure of a host of dimensionally regulated Feynman integrals as well as processes in quantum chromodynamics (QCD). Since then, their presence has been confirmed in more examples of integrals~\cite{He:2021esx,He:2021fwf,He:2021non,He:2021mme,He:2021zuv,He:2021eec,He:2022ctv,He:2022tph,Zhao:2023okw} and imprints of their relevance have also been observed in finite remainders of five-particle QCD amplitudes~\cite{Badger:2021nhg,Badger:2021ega,Abreu:2021asb,Badger:2022ncb}. A review of these developments may be found in Chapter 5~\cite{Papathanasiou:2022lan} of the SAGEX review on scattering amplitudes~\cite{Travaglini:2022uwo}. 

While we will define cluster algebras in more detail in the next section, we can convey their essence with examples of finite rank-two cluster algebras, which will play a central role in what follows: These consist of \emph{cluster variables} $a_m$ for $m$ integer, grouped into unordered sets or \emph{clusters} $\{a_{m},a_{m+1}\}$, which may be obtained as rational functions of the variables of the initial cluster, $\{a_1,a_2\}$, by virtue of the \emph{mutation} operation,
\begin{equation}\label{eq:rank2mutation}
    a_{m+1}=
\begin{cases}
\tfrac{1+a_m}{a_{m-1}}&\text{if $m$ is odd}\,,\vspace{4pt} \\
\tfrac{1+a_m^L}{a_{m-1}}&\text{if $m$ is even}\,,\\
\end{cases}
\end{equation}   
where $L$ is a positive integer with $L\le 3$. The three inequivalent cases $L=1,2,3$ correspond to the $A_2$, $C_2$ and $G_2$ cluster algebras respectively, echoing the Cartan classification of semi-simple Lie groups of the same rank. For these cases, it's easy to show that $a_{i+4+2^{L-1}}=a_i$, in other words there exist a finite number of nontrivial clusters and cluster variables. For example, in the $C_2$ case the latter are,
\begin{equation}\label{eq:C2variables}
\Phi_{C_2}=\{a_1,\ldots,a_6\}=\left\{a_1,a_2,\frac{1+a_2^2}{a_1},\frac{1+a_1+a_2^2}{a_1 a_2},\frac{1+2 a_1+a_1^2+a_2^2}{a_1 a_2^2},\frac{1+a_1}{a_2}\right\}\,. 
\end{equation}
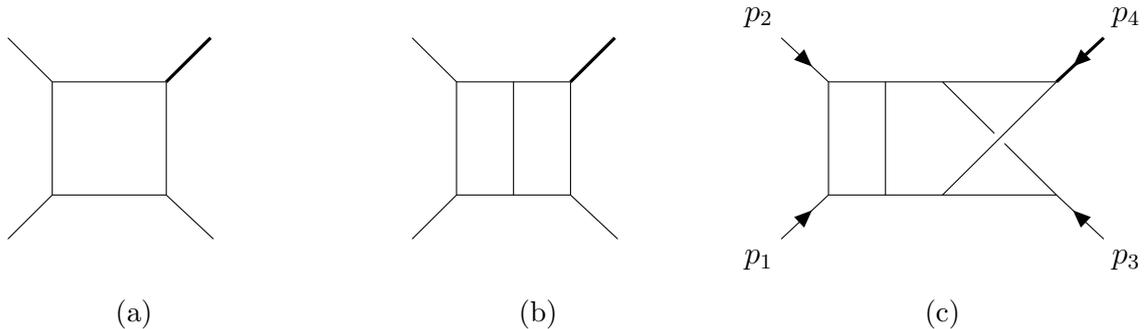
\begin{figure}
\begin{subfigure}{0.32\linewidth}
\centerline{\begin{tikzpicture}
    \begin{feynman}
    \vertex (a);
    \vertex [right =1.5cm of a] (b);
    \vertex [below =1.5cm of b] (c);
    \vertex [below =1.5cm of a] (d);
    \vertex [above left =2em of a] (x);
    \vertex [above right =2em of b] (y);
    \vertex [below right =2em of c] (z){\phantom{\(p_1\)}};
    \vertex [below left =2em of d] (w);
    \diagram*{
        (x)--(a), (y)--[very thick](b), (z)--(c), (w)--(d), (a)  -- (b)  -- (c) -- (d) -- (a),    
    }; 
    \end{feynman}
    \end{tikzpicture}}
\caption{}\label{fig:1L}
\end{subfigure}
\begin{subfigure}{0.32\linewidth}
\centerline{\begin{tikzpicture}
    \begin{feynman}
    \vertex (a);
    \vertex [right =0.75cm of a] (bmid);
    \vertex [right =0.75cm of bmid] (b);
    \vertex [below =1.5cm of b] (c);
    \vertex [below =1.5cm of a] (d);
    \vertex [right =0.75cm of d] (dmid);
    \vertex [above left =2em of a] (x);
    \vertex [above right =2em of b] (y);
    \vertex [below right =2em of c] (z){\phantom{\(p_1\)}};
    \vertex [below left =2em of d] (w);
    \diagram*{
        (x)--(a), (y)--[very thick](b), (z)--(c), (w)--(d), (a)  -- (b)  -- (c) -- (d) -- (a), (bmid) -- (dmid),    
    }; 
    \end{feynman}
    \end{tikzpicture}}
    \caption{}\label{fig:2L}
\end{subfigure}
\begin{subfigure}{0.32\linewidth}
\centerline{\begin{tikzpicture}
    \begin{feynman}
    \vertex (a);
    \vertex [right =0.75cm of a] (bleft);
    \vertex [right =0.75cm of bleft] (bmid);
    \vertex [right =1.5cm of bmid] (b);
    \vertex [below = of b] (c);
    \vertex [below = of a] (d);
    \vertex [right =0.75cm of d] (dleft);
    \vertex [right =0.75cm of dleft] (dmid);
    \vertex [above left =2em of a] (x){\(p_2\)};
    \vertex [above right =2em of b] (y){\(p_4\)};
    \vertex [below right =2em of c] (z){\(p_3\)};
    \vertex [below left =2em of d] (w){\(p_1\)};
    \diagram*{
        (x)--[fermion](a), (y)--[fermion,very thick](b), (z)--[fermion](c), (w)--[fermion](d), (a)  -- (b),   (c) -- (d) -- (a), (dleft) -- (bleft),   
    }; 
    \end{feynman}
    \draw [name path=one, opacity=0] (dmid) -- (b);
    \draw [name path=two] (bmid) -- (c);
    \path [name intersections={of=one and two,by=inter}];
    \filldraw [white] (inter) circle (2.5pt);
    \draw (dmid) -- (b);
    \end{tikzpicture}}
    \caption{}\label{fig:3L}
\end{subfigure}
\caption[]{Examples of known integrals with one off-shell leg, $p_4^2\ne 0$, at (a) one (b) two and (c) three loops. They are described by the $A_2$, $C_2$ and $G_2$ cluster algebras, respectively, with the latter case proven in this work.}\label{fig:4p1mIntegrals}
\end{figure}

The above $C_2$ cluster algebra was found to describe one of the most prominent classes of Feynman integrals studied in~\cite{Chicherin:2020umh}, which enter the calculation of amplitudes for Higgs boson plus jet production from proton-proton
collisions~\cite{Gehrmann:2011aa,Duhr:2012fh,Gehrmann:2023etk} in the heavy-top limit of QCD~\cite{Ellis:1975ap,Shifman:1979eb,Inami:1982xt}: These were all the known four-point integrals with one off-shell (or equivalently massive) leg, including the complete set of two-loop master integrals~\cite{Gehrmann:2000zt,Gehrmann:2001ck} as well as $L$-loop ladders~\cite{DiVita:2014pza,Panzer:2015ida}; see figure~\ref{fig:2L} for a representative example. More precisely, all these integrals where shown to obey differential equations of the form~\eqref{eq:CDE_intro}, where the letters $\alpha_i$ coincide (in certain kinematic parametrisation) with the $C_2$ cluster variables $a_i$ of eq.~\eqref{eq:C2variables}! More three-loop planar integral families confirming the same cluster algebraic structure were also later computed in~\cite{Canko:2021hvh,Canko:2021xmn}.

Despite this encouraging evidence, a more recent calculation~\cite{Henn:2023vbd}, see also~\cite{Syrrakos:2023mor}, has established that the $C_2$ alphabet~\eqref{eq:C2variables} is in fact too small to describe all integrals with the same kinematics at three loops: In particular, it was found that the three-loop nonplanar ladder integral family depicted in figure~\ref{fig:3L} additionally depends on two novel letters. Does this result cast doubt on the applicability of cluster algebras for Feynman integrals? The main contribution of this work is to demonstrate that  \emph{the resulting alphabet in fact corresponds to a $G_2$ cluster algebra}. We find it remarkable that the alphabets of type $A_2$, $C_2$, $G_2$ become relevant at one, two and three loops, respectively.\footnote{Note that the space of polylogarithmic functions with an $A_2$ alphabet is contained in the space of functions with a $C_2$ alphabet, and the latter is in turn contained in the space of functions with a $G_2$ alphabet.} In other words, the parameter $L$ in eq.\eqref{eq:rank2mutation} really seems to be the loop order!

In addition, we study \emph{adjacency restrictions} of the form,
\begin{equation}\label{eq:adjacencies_intro}
    \mathbf{A}_{i} \cdot \mathbf{A}_{j}=0,
\end{equation}
for some $i,j$, of the constant matrices appearing in the differential equations~\eqref{eq:CDE_intro}, for the $G_2$-alphabet integral family of figure~\ref{fig:3L}. In the realm of planar $\cN=4$ SYM theory, adjacency relations appear to encode how the cluster variables arrange themselves into the clusters~\cite{Drummond:2017ssj,Drummond:2018dfd}. They can be in essence physically interpreted as a generalisation~\cite{Caron-Huot:2019bsq} of the Steinmann relations~\cite{Steinmann,Steinmann2,Cahill:1973qp} governing the discontinuities of Feynman integrals and scattering amplitudes, and most importantly, they greatly facilitate bootstrapping the latter by drastically reducing the size of the function space containing them.

For the nonplanar triple ladder integral of figure~\ref{fig:3L}, we find 20 inequivalent adjacency relations of the form~\eqref{eq:adjacencies_intro} after transforming the original alphabet of ref.~\cite{Henn:2023vbd} to the $G_2$ cluster variables obtained from eq.~\eqref{eq:rank2mutation}, whereas before the transformation a subset of 16 of these relations were visible. In other words, the cluster algebra connection we find exposes new adjacency relations, and we showcase how these restrict the allowed function space. While these relations are not in one-to-one correspondence with how the $G_2$ cluster variables are distributed among clusters, we will also show that many of them can be understood by embedding $G_2$ inside the larger $B_3$ or $D_4$ cluster algebras.

The rest of this paper is organised as follows. In section~\ref{sec:CABasics} we briefly introduce cluster algebras and their finite type classification, before turning our attention to the rank-two cases, which will be at the heart of this work. We also review how these can be embedded into larger cluster algebras with the process of folding, and define the notion of an embedded neighbour set, which will play an important role when analysing adjacency restrictions. 

Then, in section~\ref{sec:Alphabet} we move on to discuss the alphabet containing all currently computed four-point one-mass integrals through three loops, and, after reviewing certain general procedures for proving the equivalence of alphabets, we apply them to demonstrate that the alphabet in question is described by a $G_2$ cluster algebra. Section~\ref{sec:adjacency} is dedicated to the study of adjacency relations for the single topology with novel letters, the nonplanar triple ladder. We present the adjacency restrictions we observe for the integral in question, we explain them to a great extent with the help of embeddings to larger cluster algebras, and we demonstrate that they indeed lead to a significant reduction in the size of the relevant polylogarithmic function space. Finally, section~\ref{sec:OutlookConclusions} we present our conclusions and discuss open questions for the future.

\section{Basics of Cluster Algebras}\label{sec:CABasics}
\subsection{Definitions and finite type classification}\label{subsec:CAdefinitions}
Here we give a brief introduction on the basics of cluster algebras. This relatively new branch of contemporary mathematics, was originally  motivated by the study of  representation theory and the study of quantum groups \cite{1021.16017,1054.17024,CAIII,CAIV}. However, it appeared to be very useful in many other directions both in terms of mathematics and physics \cite{Felikson_2012,keller2012cluster,Feng:2001bn,kontsevich2008stability}. For more details on its mathematical foundations one can follow \cite{Fomin2016,Fomin2017}.
\paragraph{}
For any positive $n$, a cluster algebra $\mathcal{A}$ of rank $n$ is a commutative ring with unit and no zero divisors. The structure includes a distinguished set of generators $\mathbf{a} \coloneqq \{a_1,a_2,\ldots,a_n\}$, called \textit{cluster variables}, which group into overlapping subsets called \textit{clusters}. The cardinality of each cluster is equal to the \textit{rank} of the cluster algebra.
The clusters and the cluster variables are built constructively. One starts from an initial cluster and builds the rest through an operation called \textit{mutation}. The mutation rule is provided by a skew-symmetrisable integer valued $n \times n$ matrix $B$ called the \textit{exchange matrix} with components $b_{ij}$. Any pair of  data $(\mathbf{a},B)$ is called a  \textit{seed}. Mutation leads to new cluster variables and new exchange matrices. Mutation can be performed on any cluster variable and transforms the whole seed. More precisely, mutating $(\mathbf{a},B)$ on the $k$-th variable $(1 \leq k \leq n)$ we obtain a new seed ($\mathbf{a^{'}},B^{'}$) with 
\begin{equation} \label{eq:matrixmutation}
b^{\prime}_{ij}=\begin{cases}
    -b_{ij} & \text{ for } i=k \text{ or } j=k,   \\
    b_{ij}+[-b_{ik}]_{+}b_{kj}+b_{ik}[b_{kj}]_{+}  &  \text{ otherwise, }
\end{cases}
\end{equation}
where we denoted $[b_{ij}]_{+}=\max(0,b_{ij})$. Moreover, the cluster variables mutate according to 
\begin{equation}\label{eq:variablemutation}
a^{\prime}_i=
\begin{cases}
   a_{k}^{-1} \left(\prod_{i=1}^{{n}}a_{i}^{[b_{ik}]_{+}}+\prod_{i=1}^{n}a_{i}^{[-b_{ik}]_{+}}\right) & \text{ if }  i=k \\
    a_{i} & \text{ if }  i \neq k
\end{cases}
\end{equation}
Consecutive mutations, in principle,  produce new variables on each iteration. Note that in any case the number of cluster variables remains the same. When this procedure ends, namely when the cluster variables constitute a finite set, the cluster algebra is said to be of \textit{finite type}. On the contrary, cluster algebras with infinite distinct cluster variables are said to be of \textit{infinite type}. 

In this work we will be interested in finite cluster algebras, whose classification is identical to that of semisimple Lie algebras. To this end, one associates a symmetrisable generalised Cartan matrix $A(B)$ to the skew-symmetrisable exchange matrix $B$ of the cluster algebra in the following way
\begin{equation}\label{eq:cartan}
    a_{ij}=\begin{cases}
        2,& \text{ if } i=j  \\
        -|b_{ij}|,& \text{ if } i \neq j.
    \end{cases}
\end{equation}
It has been proven in \cite{1021.16017,1054.17024}, that cluster algebras are finite if and only if they contain an exchange matrix $B$ such that $A(B)$ is a Cartan matrix of finite type. Therefore, the classification of finite cluster algebras amounts to the classification of finite Cartan matrices into types $A_n,B_n,C_n,D_n,E_6,E_7,F_4,G_2$. 

Before moving on to our main example of rank-two finite cluster algebras, let us define two more concepts we will make use of in what follows. The content of a cluster algebra can be visualised in its \emph{exchange graph}, where each cluster is represented as a vertex, and each mutation from a cluster to another as an edge between them. As an explicit example, the exchange graph of the $G_2$ cluster algebra may be found in figure~\ref{fig:exchG2} below. Finally, the \emph{neighbour set} of a cluster variable is the set of all cluster variables appearing with the latter in some cluster.

\subsection{The rank-two cluster algebras $A_2, C_2, G_2$}\label{subsec:Rank2CA}
As stated in the introduction, our main focus will be on the rank two cluster algebras of types $A_2,C_2,G_2$. These can be obtained by an initial seed with cluster variables $\{a_1,a_2\}$ and exchange matrices given by
\begin{equation}\label{eq:exchrank2}
B=\begin{pmatrix}
0 && 1\\
-L && 0 \\
\end{pmatrix},
\end{equation}
where $L=1,2,3$ corresponds to the $A_2,C_2,G_2$ case, respectively. Indeed, by computing the generalised Cartan matrix of eq.~\eqref{eq:cartan},
\begin{equation}\label{eq:cartanrank2}
A(B)=\begin{pmatrix}
2 && -1 \\
-L && 2
\end{pmatrix},
\end{equation}
we can recognise that it corresponds to the Cartan matrix of the aforementioned rank-two Lie algebras.

Applying the mutation rule of eq.~\eqref{eq:matrixmutation} to the exchange matrix~\eqref{eq:exchrank2}, it is easy to check that it only switches sign. Since the corresponding mutation rule for the cluster variables, eq.~\eqref{eq:variablemutation}, is invariant under this sign change, the latter equation then simplifies to the form~\eqref{eq:rank2mutation} we stated in the introduction. 

As the cluster algebras in question are finite, starting with the initial cluster $\{a_1,a_2\}$ and performing mutations generates a finite number of distinct clusters and variables, after which we land back to the initial cluster. In the $L=2$ or $C_2$ case the collection of distinct cluster variables from all clusters yields the set quoted in eq.~\eqref{eq:C2variables}, and similarly in the $L=3$ or $G_2$ case we obtain,

\begin{align}\label{eq:G2alphabet}
    \Phi_{G_2}= 
    & \{{\color{cyan}{a_1}},{\color{cyan}{a_2}},{\color{cyan}{\frac{1+a_2^3}{a_1}}},{\color{cyan}{\frac{1+a_1+a_2^3}{a_1 a_2}}},{\color{violet}{\frac{1+a_1^3+3 a_1^2+3 a_1 a_2^3+3 a_1+a_2^6+2 a_2^3}{a_1^2 a_2^3}}},{{\color{blue}{\frac{1+a_1^2+2 a_1+a_2^3}{a_1 a_2^2}}}},
    \nonumber\\
    &\quad {\color{violet}{\frac{1+a_1^3+3 a_1^2+3 a_1+a_2^3}{a_1 a_2^3}}},{\color{cyan}{\frac{1+a_1}{a_2}}}\}.
    \end{align}
The colour-coding is explained as follows: The subset of \textcolor{cyan}{cyan} variables is multiplicatively equivalent to the $\textcolor{cyan}{A_2}$ cluster variables after the replacement
\be
a_2\to a_2^{1/3}\,,
\ee
or in other words the logarithms of the variables span the same linear space.\footnote{As will be elaborated on below, polylogarithmic function spaces with cluster variables as letters are defined up to this equivalence, hence our interest in it.} Similarly, by adding the \textcolor{blue}{blue} variable to them one obtains a subset that is multiplicatively equivalent to the $\textcolor{blue}{C_2}$ cluster algebra after the replacement
\be\label{eq:G2toC2rep}
a_2\to a_2^{2/3}\,.
\ee
More explicitly, by dropping the purely $\textcolor{violet}{G_2}$ \textcolor{violet}{violet} variables and using \eqref{eq:G2toC2rep} in eq.~\eqref{eq:G2alphabet}, it is easy to check that the numerators of the remaining variables match (up to overall rational exponents) those of eq.~\eqref{eq:C2variables}.

\begin{figure}
\centering
\begin{tikzpicture}[scale=2,every node/.style={draw},inner sep=2pt]
   \begin{scope}[name prefix = bottom-]
    \node (A) at (0,0) {$\color{cyan}{a_1},\color{cyan}{a_2}$};
    \node (B) at (1,0) {$\color{cyan}{a_2},\color{cyan}{a_3}$};
    \node (C) at (2,-1) {$\color{cyan}{a_3},\color{cyan}{a_4}$};
    \node (D) at (2,-2) {$\color{cyan}{a_4},\color{violet}{a_5}$};
    \node (E) at (1,-3) {$\color{violet}{a_5},\color{blue}{a_6}$};
    \node (F) at (0,-3) {$\color{blue}{a_6},\color{violet}{a_7}$};
    \node (G) at (-1,-2) {$\color{violet}{a_7},\color{cyan}{a_8}$};
    \node (H) at (-1,-1) {$\color{cyan}{a_8},\color{cyan}{a_1}$};
    \draw[cyan] (A) -- (B) ;
    \draw[cyan] (B) -- (C);
    \draw[cyan] (C) -- (D);
    \draw[violet] (D) -- (E);
    \draw[blue] (E) -- (F);
    \draw[violet] (F) -- (G);
    \draw[cyan] (G) -- (H);
    \draw[cyan] (H) -- (A);
  \end{scope}
\end{tikzpicture}
\caption{Exchange graph of the $\color{violet}{G_2}$ cluster algebra, with the expressions for the letters $a_i$ given in eq.~\eqref{eq:G2alphabet} in the order they appear. Removing $\color{violet}{a_5},\color{violet}{a_7}$ (and $\color{blue}{a_6}$) and fusing the clusters containing them essentially yields the $\color{blue}{C_2}$ ($\color{cyan}{A_2}$) cluster algebra, as discussed in the text.}
\label{fig:exchG2}
\end{figure}
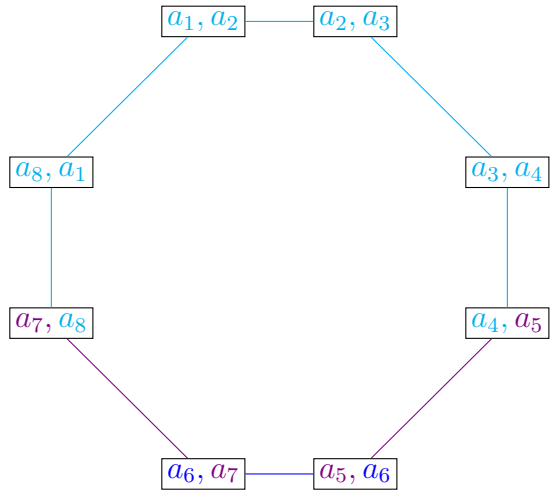

Apart from the cluster variables per se, we will also be interested in how these arrange themselves into clusters. This information is visualised in the exchange graph, as defined in the previous subsection, and which for the case of $G_2$ is depicted in figure~\ref{fig:exchG2}. The fact that the $C_2$ and $A_2$ cluster variables are contained, up to multiplicative equivalence, in $G_2$, induces a relation between the two cluster algebras also at the level of their exchange graphs: The $C_2$ exchange graph can be obtained from the $G_2$ one by dropping any purely $\color{violet}{G_2}$ variable as well as contracting the edges of the two clusters containing it, and similarly for $A_2$.

From the exchange graph we may also easily infer that the $G_2$ cluster variable $a_{i}$ may be found together with $a_{i-1}$ and $a_{i+1}$ in some cluster. This is precisely the information encoded in the notion of the neighbour set also defined at the end of the previous section, which in this case reads,
\begin{equation}\label{eq:nsG2}
ns_{G_2}(a_i)=\{a_{i-1},a_{i},a_{i+1}\}\,\quad i=1,\ldots,8.    
\end{equation}

\subsection{Folded cluster algebras and embedded neighbour sets}\label{subsec:FoldEmbed}

Every non-simply laced Lie algebra, or its corresponding Dynkin diagram, may be obtained from a simply laced one from a procedure known as \emph{folding}. In particular, we have
\be\label{eq:ADEfold}
\begin{aligned}
A_{2n-1}&\to C_n\,,& D_4&\to G_2\,,\\
D_{n+1}&\to B_n\,,& E_6 &\to F_4\,,
\end{aligned}
\ee
and the same procedure also carries over to the associated cluster algebras~\cite{Fomin:2001rc}\footnote{From the relation~\eqref{eq:cartan} between Cartan and exchange matrices, it also follows that (non-)simply laced Lie algebras are associated to skew-symmetric (skew-symmetrisable) cluster algebras.}. Conversely, this provides an embedding of a skew-symmetrisable cluster algebra inside a larger cluster algebra. We will make use of this type of embedding for $G_2$ in subsection~\ref{subsec:G2inD4}, where we will see that it has implications for the adjancency relations of the form \eqref{eq:adjacencies_intro} for the differential equations of the three-loop nonplanar ladder integral.

In the rest of this subsection, we review how to fold cluster algebras, also using the $A_3\to C_2$ example to illustrate the general process. We will also show how folding naturally extends the notion of neighbour sets to \emph{embedded neighbour sets}, see in particular Definition~\ref{def:ENS} below. The reader interested primarily in the cluster-algebraic structure  of alphabets, rather than of adjacency relations, may thus choose to directly skip to the next section.

At the level of Dynkin diagrams, folding may be thought of as an identification between different nodes that are mapped into each other by an (outer automorphism) symmetry leaving  the entire diagram invariant. At the level of a cluster algebra, this symmetry is in turn with respect to permutations of the rows of the exchange matrix;\footnote{As the exchange matrix we begin with is skew-symmetric, also its columns will respect the same symmetry.} Folding of the cluster algebra essentially amounts to equating the cluster variables associated to the rows in question, and only considering simultaneous mutations of these variables (and usual mutations of the rest)~\cite{Fomin2017}. 

In this manner, one produces the subset of the seeds of the skew-symmetric cluster algebra, whose exchange matrix will respect an analogous permutation symmetry. As a consequence, these clusters will also have the same number of cluster variables equal after the aforementioned replacement, and we may use the count of distinct variables as a criterion to select the subset in question when starting from the entire skew-symmetric cluster algebra. Furthermore, from the cluster variable mutation rule~\eqref{eq:variablemutation}, it is evident that 
\be
a_{i}^{[\pm b_{ik}]_{+}}a_{j}^{[\pm b_{jk}]_{+}}\xrightarrow{a_j=a_i} a_{i}^{[\pm b_{ik}]_{+}+[\pm b_{jk}]_{+}}\,.
\ee
In other words, the identification of two cluster variables implies that their corresponding rows in the exchange matrix need to be replaced by the sum of the two, also appropriately eliminating rows so as to end up with a rectangular matrix. All in all, from these considerations we obtain the following general folding procedure.

\smallskip
\noindent {\bf Folding cluster algebras:} 
\begin{enumerate}
    \item Start with a skew-symmetric $ADE$ cluster algebra appearing on the left of any arrow in eq.~\eqref{eq:ADEfold}, and pick a seed whose exchange matrix is symmetric under a certain permutation of its rows. 
    \item Equate the cluster variables related by this permutation, and select the subset of all clusters containing the same numbers of equal cluster variables.
    \item In each of the selected clusters, for each collection of equal cluster variables, replace the corresponding rows of the exchange matrix with a single row summing them up. Eliminate the columns of the matrix as well as the duplicate cluster variables having the same position as the eliminated rows.
    \item The seeds thus obtained are seeds of the skew-symmetrisable cluster algebra at the other end of the arrow in eq.~\eqref{eq:ADEfold}.
\end{enumerate}
Let us see this procedure at work in the example of $A_3\to C_2$ folding. We start with the $A_3$ seed with cluster variables and exchange matrix, respectively,
\begin{equation}\label{eq:A3_initialB}
\{a_0,a_1,a_2\}\,,\quad    \bar B=\begin{bmatrix}
0 & -1 & 0  \\
1 & 0 & 1  \\
0 & -1 & 0 
\end{bmatrix}\,,
\end{equation}
and for our purposes it will be sufficient to consider two more seeds of the cluster algebra,
\be
\left\{\frac{1 + a_1}{a_0}, a_1, a_2\right\}\,,\quad\bar{B}'=\left[
\begin{array}{ccc}
 0 & 1 & 0 \\
 -1 & 0 & 1 \\
 0 & -1 & 0 \\
\end{array}
\right]\,,\label{eq:A3notC2}
\ee
\be
\left\{\frac{1 + a_1}{a_0}, a_1, \frac{1 + a_1}{a_2}\right\}\,,\quad \bar{B}''=\left[
\begin{array}{ccc}
 0 & 1 & 0 \\
 -1 & 0 & -1 \\
 0 & 1 & 0 \\
\end{array}
\right]\,,
\label{eq:A3andC2}\ee
obtained by mutating first $a_0$ and then $a_2$ in~\eqref{eq:A3_initialB}. The matrix $\bar{B}$ in the latter formula is symmetric under  the exchange of the first and third row and column, so according to step 2 above we may set $a_0=a_2$. This renders equal two cluster variables not only in this seed but also in the seed of eq.~\eqref{eq:A3andC2}, and so we select these seeds. On the contrary the three cluster variables of the seed~\eqref{eq:A3notC2} remain distinct and so we discard it (also notice that, unlike the two seeds we selected, the exchange matrix of this seed does not have a permutation symmetry).

We now proceed to produce the $C_2$ seeds corresponding to eqs.~\eqref{eq:A3_initialB} and~\eqref{eq:A3andC2} by identifying the equal cluster variables as described in step 3 above. In both selected clusters the permutation symmetry is with respect to the first and third row, and we may choose to replace the latter with the sum of the two. This eliminates the first row, and consequently also the first column of the matrix, as well as the first cluster variable. Therefore folding eqs.~\eqref{eq:A3_initialB} and~\eqref{eq:A3andC2} yields
\begin{align}
\{a_1,a_2\}\,,&\quad      B=\begin{bmatrix}
    0 & 1 \\
    -2 & 0
    \end{bmatrix}\,,\\
\left\{a_1, \frac{1 + a_1}{a_2}\right\}\,,&\quad B''=\left[
\begin{array}{cc}
 0 & -1  \\
 2 & 0  \\
\end{array}
\right]\,,
\end{align}
respectively, and it's easy to show that both are indeed seeds of the $C_2$ cluster algebra: The first of the above equations is nothing but the $C_2$ initial seed with an exchange matrix equal to eq.~\eqref{eq:exchrank2} with $L=2$, whereas the second equation the seed obtained by mutating $a_2$ in the former.

Similarly, carrying out this procedure for all 14 of the $A_3$ seeds selects 6 of them, which are found to be equal to the $C_2$ seeds as expected by the coarsening of the $G_2$ exchange graph shown in figure~\ref{fig:exchG2}. It also follows that the set of cluster variables of the folded cluster algebra is directly obtained from that of the unfolded cluster algebra we begin with, after the variable identification discussed in step 2 above. For the $A_3$ example at hand, it is in particular simple to check that its set of cluster variables,
\begin{equation*}
   \left\{a_0,a_1,a_2,\tfrac{1+a_1}{a_0},\tfrac{1+a_0 a_2}{a_1},\tfrac{1+a_1}{a_2},\tfrac{1+a_1+a_0 a_2}{a_0 a_1},\tfrac{1+a_1+a_0 a_2}{a_1 a_2},\tfrac{1+2 a_1+a_1^2+a_0 a_2}{a_0 a_1 a_2}\right\}\,,
\end{equation*}
indeed reduces to that of $C_2$, eq.~\eqref{eq:C2variables}, after the $A_3\to C_2$ folding replacement $a_0=a_2$ and the elimination of duplicate elements.

Before concluding this section, let us present a final related concept that will be very useful when analysing adjacency relations in section~\ref{sec:adjacency}. As the folding procedure we have described may be thought of as an embedding of a skew-symmetrisable cluster algebra inside a larger cluster algebra, it also allows us to generalise the notion of the neighbour set defined at the end of subsection~\ref{subsec:CAdefinitions}, and also illustrated in the example of the $G_2$ cluster algebra in eq.~\eqref{eq:nsG2}. Namely, cluster variables that do not appear together in clusters of the folded cluster algebra, may appear together in clusters of the larger cluster algebra containing it. This gives rise to the notion of an \emph{embedded neighbour set}, which may be formally defined as follows.

\begin{definition}[Embedded neighbour set]\label{def:ENS}
Let $a_i\in A$ and $f_i \in F$ be cluster variables of two cluster algebras related by folding $A\to F$, which in particular equates $a_j=f_i$ for some indices $j$ and all indices $i$. Then the embedded neighbour set of $F\subset A$ is given by
\be
ns_{F\subset A}(f_i)=\bigcup_j ns_A(a_j)\Big|_{a_j=f_i}\,.
\ee
\end{definition}
In other words, we first compute the neighbour sets for the variables of the cluster algebra $A$, and then set some of them equal, as dictated by folding. For the subset of variables that become equal to the cluster variables of $F$ after this replacement, we take the union of their neighbour sets, where the same replacement is also applied.

To make the embedded neighbour set definition more transparent, we apply it to our $A_3\to C_2$ example. The $A_3$ neighbour sets are,
\begin{align}
ns_{A_3}(a_0)&=\left\{a_0,a_1,a_2,\frac{1+a_1}{a_2},\frac{1+a_0 a_2}{a_1},\frac{1+a_1+a_0 a_2}{a_1 a_2}\right\}\,,\nonumber\\
ns_{A_3}(a_1)&=\left\{a_0,a_1,a_2,\frac{1+a_1}{a_0},\frac{1+a_1}{a_2}\right\}\,,\nonumber\\
ns_{A_3}(a_2)&=\left\{a_0,a_1,\frac{1+a_1}{a_0},a_2,\frac{1+a_0 a_2}{a_1},\frac{1+a_1+a_0 a_2}{a_0 a_1}\right\}\,,\nonumber\\
ns_{A_3}\left(\frac{1+a_1}{a_0}\right)&=\left\{a_1,\frac{1+a_1}{a_0},\frac{1+a_1}{a_2},a_2,\frac{1+a_1+a_0 a_2}{a_0 a_1},\frac{1+2 a_1+a_1^2+a_0 a_2}{a_0 a_1 a_2}\right\}\,,\nonumber\\
ns_{A_3}\left(\frac{1+a_0 a_2}{a_1}\right)&=\left\{a_0,a_2,\frac{1+a_0 a_2}{a_1},\frac{1+a_1+a_0 a_2}{a_0 a_1},\frac{1+a_1+a_0 a_2}{a_1 a_2}\right\}\,,\\
ns_{A_3}\left(\frac{1+a_1}{a_2}\right)&=\left\{a_0,a_1,\frac{1+a_1}{a_0},\frac{1+a_1}{a_2},\frac{1+a_1+a_0 a_2}{a_1 a_2},\frac{1+2 a_1+a_1^2+a_0 a_2}{a_0 a_1 a_2}\right\}\,,\nonumber
\end{align}
\begin{align}
ns_{A_3}\left(\textstyle\frac{1+a_1+a_0 a_2}{a_0 a_1}\right)&=\left\{\frac{1+a_1}{a_0},a_2,\frac{1+a_0 a_2}{a_1},\frac{1+a_1+a_0 a_2}{a_0 a_1},\frac{1+a_1+a_0 a_2}{a_1 a_2},\textstyle\frac{1+2 a_1+a_1^2+a_0 a_2}{a_0 a_1 a_2}\right\}\,,\nonumber\\
ns_{A_3}\left(\textstyle\frac{1+2 a_1+a_1^2+a_0 a_2}{a_0 a_1 a_2}\right)&=\left\{\frac{1+a_1}{a_0},\frac{1+a_1}{a_2},\frac{1+a_1+a_0 a_2}{a_0 a_1},\frac{1+a_1+a_0 a_2}{a_1 a_2},\textstyle\frac{1+2 a_1+a_1^2+a_0 a_2}{a_0 a_1 a_2}\right\}\,,\nonumber\\
ns_{A_3}\left(\textstyle\frac{1+a_1+a_0 a_2}{a_2 a_1}\right)&=\left\{a_0,\frac{1+a_1}{a_2},\frac{1+a_0 a_2}{a_1},\frac{1+a_1+a_0 a_2}{a_0 a_1},\frac{1+a_1+a_0 a_2}{a_1 a_2},\textstyle\frac{1+2 a_1+a_1^2+a_0 a_2}{a_0 a_1 a_2}\right\}\,.\nonumber
\end{align}
Upon the identification $a_0=a_2$ that performs the $A_3\to C_2$ folding, two more pairs of cluster variables become equal to each other (those appearing on the left-hand side of lines 4-7 and 5-9 above). We observe that the neighbour sets of the cluster variables that are identified with each other also coincide, and altogether the $C_2\subset A_3$ embedded neighbour sets thus become,
\begin{align}
ns_{C_2\subset A_3}(a_1)=&\left\{a_1,a_2,\frac{1+a_1}{a_2}\right\}\nonumber\\
ns_{C_2\subset A_3}(a_2)=&\left\{a_1,a_2,\frac{1+a_1}{a_2},\frac{1+a_2^2}{a_1},\frac{1+a_1+a_2^2}{a_1 a_2}\right\}\,,\nonumber\\
ns_{C_2\subset A_3}\left(\frac{1+a_2^2}{a_1}\right)=&\left\{a_2,\frac{1+a_2^2}{a_1},\frac{1+a_1+a_2^2}{a_1 a_2}\right\}\,,\\
ns_{C_2\subset A_3}\left(\frac{1+a_1}{a_2}\right)=&\left\{a_1,a_2,\frac{1+a_1}{a_2},\frac{1+a_1+a_2^2}{a_1 a_2},\frac{1+2 a_1+a_1^2+a_2^2}{a_1 a_2^2}\right\}\,,\nonumber\\
ns_{C_2\subset A_3}\left(\frac{1+2 a_1+a_1^2+a_2^2}{a_1 a_2^2}\right)=&\left\{\frac{1+a_1}{a_2},\frac{1+a_1+a_2^2}{a_1 a_2},\frac{1+2 a_1+a_1^2+a_2^2}{a_1 a_2^2}\right\}\,,\nonumber\\
ns_{C_2\subset A_3}\left(\frac{1+a_1+a_2^2}{a_1 a_2}\right)=&\left\{a_2,\frac{1+a_1}{a_2},\frac{1+a_2^2}{a_1},\frac{1+a_1+a_2^2}{a_1 a_2},\frac{1+2 a_1+a_1^2+a_2^2}{a_1 a_2^2}\right\}\,.\nonumber
\end{align}
Notice in particular that some of these neighbour sets are \emph{larger} than the usual $C_2$ neighbour sets of the same cluster variables. Indeed the latter always contain three elements, as every $C_2$ cluster variable appears in two clusters related by a mutation.

\section{Nonplanar 3-loop 1-mass Ladder: Alphabet}\label{sec:Alphabet}

\begin{figure}
\centerline{\begin{tikzpicture}
    \begin{feynman}
    \vertex (a);
    \vertex [right = of a] (bleft);
    \vertex [right = of bleft] (bmid);
    \vertex [right = of bmid] (b);
    \vertex [below = of b] (c);
    \vertex [below = of a] (d);
    \vertex [right = of d] (dleft);
    \vertex [right = of dleft] (dmid);
    \vertex [above left =2em of a] (x){\(p_2\)};
    \vertex [above right =2em of b] (y){\(p_4\)};
    \vertex [below right =2em of c] (z){\(p_3\)};
    \vertex [below left =2em of d] (w){\(p_1\)};
    \diagram*{
        (x)--[fermion](a), (y)--[fermion,very thick](b), (z)--[fermion](c), (w)--[fermion](d), (a)  -- (b),   (c) -- (d) -- (a), (dleft) -- (bleft),   
    }; 
    \end{feynman}
    \draw [name path=one, opacity=0] (dmid) -- (b);
    \draw [name path=two] (bmid) -- (c);
    \path [name intersections={of=one and two,by=inter}];
    \filldraw [white] (inter) circle (2.5pt);
    \draw (dmid) -- (b);
    \end{tikzpicture}
}
\caption[]{The B1 topology producing new letters.}
\label{fig:B1Topology}
\end{figure}
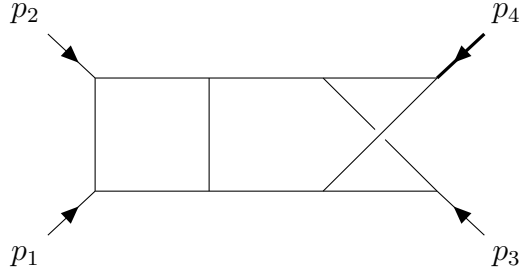

Our goal will be to understand the cluster-algebraic structure of three-loop four-point integrals with one external leg offshell/massive, and everything else massless. Many integral families with these kinematics have been recently calculated in~\cite{Canko:2021hvh,Canko:2021xmn,Henn:2023vbd,Syrrakos:2023mor}, and all of them continue to be described by a $C_2$ cluster algebra relevant at two loops~\cite{Chicherin:2020umh}, except the nonplanar topology shown in figure~\ref{fig:B1Topology}. As its alphabet contains that of all the other integrals, from now on we will thus restrict our attention to this nonplanar triple ladder.

We choose to label the external momenta as shown in the figure, with $p_1^2=p_2^2=p_3^2=0$ and $p_4^2 \neq 0$. The kinematic variables of the process are thus
 \begin{equation}\label{eq:dimful_vars}
     s=(p_1+p_2)^2, t=(p_1+p_3)^2, p_4^2,
 \end{equation}
and, following the conventions of~\cite{Henn:2023vbd}, the symbol alphabet for the canonical differential equations~\eqref{eq:CDE_intro} of all currently known integrals with these kinematics is contained in
 \begin{equation}
\begin{aligned}
    \mathbf{\Phi}_0=\{\alpha_0,\alpha_1,\ldots,\alpha_8\}=\{p_4^2,s,t,-p_4^2+s+t,-p_4^2+s,s+t,\\
    -(p_4^2-s)^2+p_4^2t,s^2-p_4^2(s-t)\}.
\end{aligned}
 \end{equation}
 Without loss of generality, we can always render the alphabet dimensionless by dividing out with one of the dimensionful letters, which is taken as the overall scale of the Feynman integral. In our case we will take this overall scale to be $\alpha_0=p_4^2$, and in particular define the dimensionless version of the kinematic variables~\eqref{eq:dimful_vars} as
\begin{equation}
    z_1=\frac{-s}{-p_4^2}, \quad z_2=\frac{-t}{-p_4^2}.
\end{equation}
The dimensionless alphabet in terms of these variables finally becomes,
\begin{equation}\label{eq:B1alphabet}
    \mathbf{\Phi}=\{{\textcolor{cyan}{z_1}},{\textcolor{cyan}{z_2}},{\textcolor{cyan}{1-z_1-z_2}},{\textcolor{cyan}{1-z_1}},{\textcolor{cyan}{1-z_2}},{\textcolor{blue}{z_1+z_2}},{\textcolor{violet}{1-2z_1+z_1^2-z_2}},{\textcolor{violet}{z_1-z_1^2-z_2}}\},
\end{equation}
where we again use colour-coding to denote the letters that first appear at \textcolor{cyan}{one}, \textcolor{blue}{two} and \textcolor{violet}{three} loops.

We would like to explore whether the alphabet~\eqref{eq:B1alphabet} has a cluster-algebraic interpretation. To this end, we will first need to review and refine the general framework for proving the equivalence of different sets of alphabets.

\subsection{General procedure for equivalence of alphabets}

Our starting point is the well-known fact that under linear transformations of the form
\be\label{eq:alphabet_transform}
\dlog\alpha_i=\sum_{j} M_{ij}\,\dlog\alpha'_j\,,
\ee
where $M_{ij}$ are the elements of a square invertible rational matrix, 
the canonical differential equations~\eqref{eq:CDE_intro} preserve their general form. Specifically, they transform to
\begin{equation}
    d\mathbf{f}(\vec{z};\epsilon)=\epsilon \left[\sum_{j}\,\mathbf{A}'_{j} \,\dlog\alpha'_j(\vec{z})\right]\mathbf{f}(\vec{z};\epsilon),
\end{equation}
where
\be\label{eq:coefficient_transform}
\mathbf{A}'_{j}=\sum_i \mathbf{A}_{i}M_{ij}\,.
\ee
Therefore, any two symbol alphabets $\{\alpha_i\}$ and $\{\alpha'_i\}$ are considered equivalent if they are related by a transformation of the form~\eqref{eq:alphabet_transform}, or in other words alphabets are only defined up to the equivalence relation $\{\alpha_i\}\sim \{\alpha'_i\}$, rather than uniquely. 

Provided the alphabets $\{\alpha_i\}$ and  $\{\alpha'_i\}$ depend on the same variables $\vec z$, one may easily check if they are equivalent in a numerical fashion, see e.g. \cite{Abreu:2020jxa}: One forms the list
\be\label{eq:ABlist}
L(\vec z)=\{\log|\alpha_1(\vec z)|,\ldots,\log|\alpha_N(\vec z)|,\log|\alpha'_1(\vec z)|,\ldots,\log|\alpha'_N(\vec z)|\}\,,
\ee
where the absolute value is placed so as to throw away any sign information which is not relevant for symbol letters, and for definitiveness we have specified the size of the two alphabets to be $N$. Next, one evaluates this list for (at least) $2N$ random values of the variables, $\vec z^{(1)},\ldots, \vec z^{(2N)}$, and from them constructs the matrix,
\be\label{eq:ABmatrix}
R=\left(
\begin{array}{c}
L(\vec z^{(1)})\\
\vdots\\
L(\vec z^{(2N)})
\end{array}
\right)\,.
\ee
Then, checking whether the two alphabets are equivalent or not boils down to the computation of the rank of this matrix,
\be\label{eq:AB_equiv_test}
\text{rank}(R)=N\quad \Leftrightarrow\quad \{\alpha_i\}\sim \{\alpha'_i\}\,.
\ee

More often, however, we are interested in comparing alphabets that depend on \emph{different variables}, $\{\alpha_i(\vec z)\}$, $\{\alpha'_i(\vec z\,'\}$. In other words, we also need to find the transformation
\be\label{eq:general_variable_transformation}
\vec z\,'= \vec g(\vec z)\,,
\ee
so as to be able to carry out the equivalence test~\eqref{eq:ABlist}-\eqref{eq:AB_equiv_test}. The key idea here is that \emph{if the variables $\vec z$ are themselves letters, then their general form is also constrained by eq.~\eqref{eq:alphabet_transform}}.\footnote{The requirement that the variables are themselves letters is not a limitation in practice, as we can always pick a subset of algebraically independent letters as variables.} If this requirement holds, and $\{\alpha_i\}\sim \{\alpha'_i\}$, then taking the exponential of the latter relation for the variables $\vec z$ yields,
\be\label{eq:variable_transform}
z'_m \in \{\alpha_i'\}\quad \Rightarrow\quad  z'_m=c_m \prod_{j=1}^N \alpha_j(\vec z)^{n_{mj}}\,,\,\,m=1,\ldots,d\,,
\ee
where for concreteness we have assumed that the number of independent variables of both alphabets is $d$, $\vec z=(z_1,\ldots, z_d)$ and similarly for the primed case.
 
The above formula equips us with a systematic way to establish the equivalence of two alphabets $\{\alpha_i(\vec z)\}$, $\{\alpha'_i(\vec z\,')\}$, both having size $N$ and $d$ independent variables: We perform the transformation~\eqref{eq:variable_transform} on the alphabet $\{\alpha'_i(\vec z\,')\}$ for a range of different values $c_m$, $n_{mj}$, construct the matrix $R$ as in eqs.~\eqref{eq:ABlist}-\eqref{eq:ABmatrix} and compute its rank. Per eq.~\eqref{eq:AB_equiv_test}, if for certain values of $c_m$, $n_{mj}$ this rank equals $N$, then these determine the transformation~\eqref{eq:general_variable_transformation}, and prove the equivalence of the two alphabets.

We may also drastically reduce the number of matrix rank evaluations of the method, and thereby achieve a corresponding improvement in its efficiency, as follows: Assuming that we scan a range of $r$ values for each $n_{mj}$, if we were to transform all $z_m$ similtaneously the number of evaluations would be proportinal to $r^{d\cdot N}$. Instead, we \emph{split the above computation into $d$ smaller computations, one for each individual value of $m$ in eq.~\eqref{eq:variable_transform}, where we also keep only the subset of single-variable letters $\{\alpha'_{i_1}(z'_m),\ldots,\alpha'_{i_K}(z'_m) \}$ in the second half of the list~\eqref{eq:ABlist}}. Each smaller computation will involve matrices  $R$ of size $N+K$ instead of $2N$, whose rank is thus also faster to evaluate, and the number of matrix rank evaluations of all of them together will now be proportional to $d\cdot r^{N}$ instead of $r^{d\cdot N}$. Finally, we form the complete list of the two alphabets $L$ in eq.~\eqref{eq:ABlist}, and carry out the test~\eqref{eq:AB_equiv_test}, but now we scan only over the subset of $c_m, n_{mj}$ values that have already passed the test in each of the smaller computations. Usually there's just a handful of such values, so this final step has negligible computational overhead.

Of course, if a variable transformation and alphabet equivalence is discovered numerically as described above, in the end it may also be checked analytically to further confirm its correctness. Input parameters of the method we have described include the ordering of the two alphabets, the range of values we scan for $c_m$, $n_{mj}$, the range and type of numbers (integer/rational/real) for the random kinematic points, as well as the numerical precision of the rank evaluation. Finally, it is worth noting that this method can also be straightforwardly generalised so as to also look for the inclusion of one alphabet inside the other. 

\subsection{Application: Ladder alphabet = $G_2$ cluster algebra}

We now move on to apply the general procedure we have described in the previous section, in order to investigate whether the alphabet~\eqref{eq:B1alphabet} of the nonplanar triple ladder integral is equivalent to the $G_2$ cluster algebra alphabet~\eqref{eq:G2alphabet}: Both have 8 letters and depend on two variables, which are necessary conditions for their equivalence. Furthermore, the variables $z_1,z_2$ are also letters of the former alphabet, such that our method can be applied directly.

For simplicity, instead of the $G_2$ alphabet~\eqref{eq:G2alphabet} we choose the alphabet of its irreducible factors (the two are equivalent by virtue of eq.~\eqref{eq:alphabet_transform}) as our $\{\alpha'_i(a_1,a_2^3)\}$. Notice, in particular, that we pick $a_2^3$ instead of $a_2$ as a variable, since only the former appears in the irreducible factors. As the range of values for our scan we pick $c_m=\pm 1$ and $n_{mj}=\{-1,0,1\}$. Finally, we choose integer random kinematic points, and compute the rank with 100 digits of precision. 

In this manner, we find the transformation
\begin{equation}
\label{eq:transform}
\begin{aligned}
    z_1= & \frac{(1 + a_1) (1 + a_1 + a_2^3)}{a_1 a_2^3}\,, \\
   z_2= & -\frac{1 + a_1}{a_2^3}\,,
\end{aligned}
\end{equation}
and we also check analytically that it indeed proves the equivalence of the alphabets~\eqref{eq:B1alphabet} and \eqref{eq:G2alphabet}. Concretely, the letters of the former, nonplanar triple ladder alphabet in terms of the $G_2$ alphabet read,
\be\label{eq:B1toG2_alphabet}
   \Phi= \{\frac{a_4a_8}{a_2},-\frac{a_8}{a_2^2},-\frac{a_4}{a_2^2},-\frac{a_6}{a_2},\frac{a_1a_4}{a_2^2},\frac{a_3a_8}{a_2^2},\frac{a_7a_4}{a_2^2},-\frac{a_5a_8}{a_2^2}\}.
\ee

\section{Nonplanar 3-loop 1-mass Ladder: Adjacencies}\label{sec:adjacency}
\subsection{Observed adjacency restrictions}
In the previous section, we showed that the alphabet~\eqref{eq:B1alphabet}, controlling the three-loop one-mass nonplanar ladder integral, is equivalent to the one of eq.\eqref{eq:G2alphabet}, dictated by the $G_2$ cluster algebra. That is, the former alphabet is related to the latter by a transformation of the form~\eqref{eq:alphabet_transform}, and it is also interesting to investigate adjacency restrictions of the transformed coefficient matrices $\mathbf{A}'_{i}$ of this integral, eq.~\eqref{eq:coefficient_transform}, in the $G_2$ alphabet.

We find that
\begin{equation}\label{eq:observed_adjacencies}
    \mathbf{A}'_{i} \cdot \mathbf{A}'_{j}=\mathbf{A}'_{j} \cdot \mathbf{A}'_{i}=0\,\,\, \text{for} \begin{cases}
     i,j\in \{1,3,5,7\}\,\text{with } i> j\,,\\
     \\
     j=i+3\,,i=3,\ldots,6\,\text{ with } j\sim j-8\,,
    \end{cases}\,,
\end{equation}
namely we obtain 12+8=20 adjacency restrictions. These contain the 6 restrictions observed previously for the $C_2$ subalphabet of $G_2$~\cite{Chicherin:2020umh}, see also \cite{Dixon:2020bbt}, plus another 14 relations involving either or both of the new,  purely $G_2$ letters  ${\color{violet}{a_5}}, {\color{violet}{a_7}}$ in eq.~\eqref{eq:G2alphabet}.

We note that 16 of these restrictions were visible in the original alphabet~\eqref{eq:B1alphabet}. Namely our cluster-algebraic analysis already has the benefit of revealing new adjacency restrictions, and in subsection~\ref{subsec:function_counts} we will explore how such adjacency restrictions reduce the size of the relevant function space. To see how the new adjacency restrictions look like in the original alphabet~\eqref{eq:B1alphabet} of the nonplanar 3-loop 1-mass ladder, we first reexpress the relation~\eqref{eq:B1toG2_alphabet} between these letters and those of the $G_2$ alphabet in the form~\eqref{eq:alphabet_transform}, and read off the corresponding transformation matrix $M$. Plugging this in eq.~\eqref{eq:coefficient_transform}, we find that the primed coefficient matrices in the $G_2$ alphabet are related to those of the original reference~\cite{Henn:2023vbd} by,
\begin{align}
\mathbf{A}'_1=\mathbf{A}_5\,,\quad \mathbf{A}'_3=\mathbf{A}_6\,,\quad \mathbf{A}'_5=\mathbf{A}_8,\,,\quad \mathbf{A}'_6=\mathbf{A}_4\,,\quad \mathbf{A}'_7=\mathbf{A}_7\label{eq:Ap2AL1}\\
\mathbf{A}_2= -\mathbf{A}_1 - 2\mathbf{A}_2 - 2\mathbf{A}_3 - \mathbf{A}_4 - 2\mathbf{A}_5 - 2\mathbf{A}_6 - 2\mathbf{A}_7 - 2\mathbf{A}_8\,, \\
\mathbf{A}'_4=\mathbf{A}_1 + \mathbf{A}_3 + \mathbf{A}_5 + \mathbf{A}_7\,,\quad \mathbf{A}'_8=\mathbf{A}_1 + \mathbf{A}_2 + \mathbf{A}_6 + \mathbf{A}_8\,.\label{eq:Ap2AL3}
\end{align}
Clearly, the subset of adjacency restrictions~\eqref{eq:observed_adjacencies} where only matrices of eq.~\eqref{eq:Ap2AL1} participate, have precisely the same simple form in the original alphabet up to index relabellings, and correspond to the 16 visible relations we mentioned at the beginning of this paragraph. However the restrictions involving the matrices of eq.~\eqref{eq:Ap2AL3} become more complicated,
\be
\begin{aligned}
(\mathbf{A}_1 + \mathbf{A}_3 + \mathbf{A}_5 + \mathbf{A}_7) \cdot \mathbf{A}_7 =  \mathbf{A}_7 \cdot (\mathbf{A}_1 + \mathbf{A}_3 + \mathbf{A}_5 + \mathbf{A}_7) = 0\,, \\
\mathbf{A}_8 \cdot (\mathbf{A}_1 + \mathbf{A}_2 + \mathbf{A}_6 + \mathbf{A}_8) = (\mathbf{A}_1 + \mathbf{A}_2 + \mathbf{A}_6 + \mathbf{A}_8)\cdot\mathbf{A}_8    = 0\,,
\end{aligned}
\ee
and may thus only be found by simple inspection after our variable transformation, illustrating its advantage.

In $\cN=4$ SYM theory, adjacency restrictions have been observed to be in 1-1 correspondence with the distribution of variables within clusters, encoded in their neighbour sets. In the language of canonical differential equations, this `cluster adjacency'~\cite{Drummond:2017ssj} correspondence could be stated as,
\be\label{eq:N=4_adjacency}
\mathbf{A}_{i} \cdot \mathbf{A}_{j}=\mathbf{A}_{j} \cdot \mathbf{A}_{i}=0\quad\Leftrightarrow \quad \begin{aligned}
&\nexists \,\,\text{cluster containing } \alpha_i,\alpha_j\,,\\
&\text{or equivalently } \alpha_i\not\in ns(\alpha_j)\,.
\end{aligned}
\ee
Note that for such cluster adjacency restrictions the order of the matrix product doesn't matter, since all statements on the right-hand side of the double arrow are independent of the order. Another way to say this, is that $\alpha_i\not\in ns(\alpha_j)\Leftrightarrow \alpha_j\not\in ns(\alpha_i)$.

This independence from the order of the matrix product is certainly something that our observed adjacencies~\eqref{eq:observed_adjacencies} respect. But do they precisely match the $\cN=4$ cluster adjacency predictions~\eqref{eq:N=4_adjacency}? To answer this, we need to apply it to the $G_2$ neighbour sets, which we have presented in eq.~\eqref{eq:nsG2}. Since these contain three consecutive $G_2$ cluster variables, it follows that if $\cN=4$ cluster adjacency were to hold, it would imply
\be\label{eq:N=4_G2_adjacency}
\mathbf{A}'_{i} \cdot \mathbf{A}'_{j}\overset{?}{=}0\,\,\text{ for } j\neq i,i\pm 1\,,
\ee
namely 16+16+8=40 relations for $j=i\pm 2$, $j=i\pm 3$ and $j=i+4$, respectively.

Hence the $\cN=4$ cluster adjacency predictions~\eqref{eq:N=4_G2_adjacency} don't match our observed adjacencies~\eqref{eq:observed_adjacencies}, though the latter are certainly contained in the former. A similar situation was also observed for the two-loop integrals for the same kinematics~\cite{Chicherin:2020umh}. We can nevertheless aim to explain these by embedding the $G_2$ cluster algebras inside larger cluster algebras with the method of folding, reviewed together with the newly introduced concept of embedded neighbour pairs in section~\ref{subsec:FoldEmbed}. The main idea is that since the larger cluster algebra contains more clusters, the neighbour sets of the $G_2$ variables inside of it will also become bigger, such that their complements, the $\cN=4$ cluster adjacency restrictions~\eqref{eq:N=4_G2_adjacency}, will reduce in number, and may approach or coincide with the observed adjacencies~\eqref{eq:observed_adjacencies}.

\subsection{Intepretation by embedding $G_2$ inside $D_4/B_3$ cluster algebras}\label{subsec:G2inD4}

\begin{figure}
    \centering
    \includegraphics[scale=0.3,trim=0.5cm 10.5cm 0.5cm 11cm]{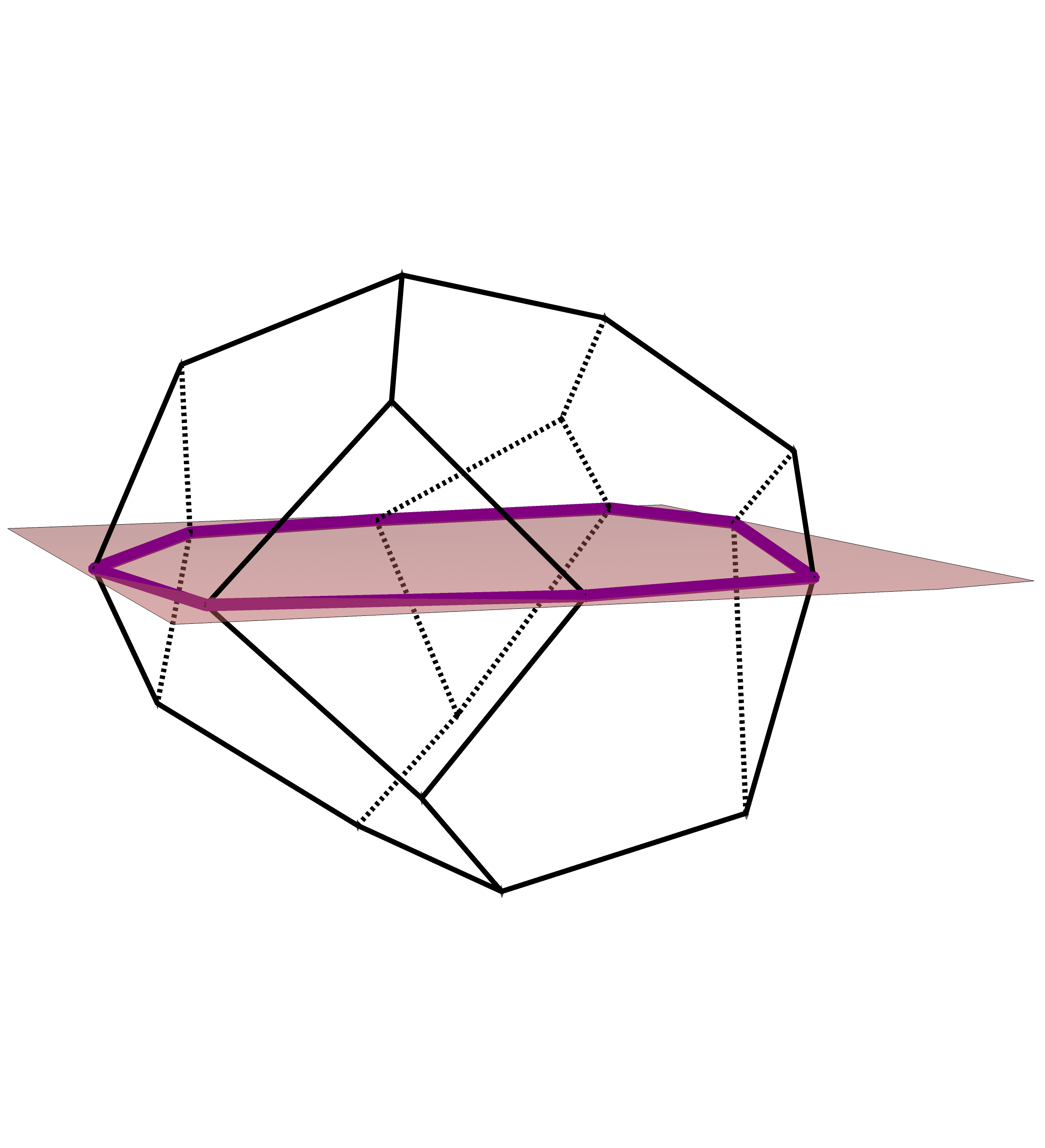}
    \caption{The three-dimensional cyclohedron with four square, four pentagonal and four hexagonal faces is the exchange graph of the $B_3$ cluster algebra. The $G_2$ exhange graph of figure~\ref{fig:exchG2} may be realised as a \textcolor{violet}{two-dimensional subspace} thereof.}
    \label{fig:exchB3}
\end{figure}

In section~\ref{subsec:FoldEmbed}, we have recalled that both the $G_2$ and the $B_3$ cluster algebras can be embedded in the larger $D_4$ cluster algebra with the method of folding. As we will explain below, due to their common ancestry, $G_2$ may also be considered as part of the $B_3$ cluster algebra, and this relation is visualised in figure~\ref{fig:exchB3}. Per Definition~\ref{def:ENS}, we find that in both $B_3$ and $D_4$ embeddings of $G_2$ the corresponding neighbour sets become,
\be\label{eq:B3D4inG2ENS}
\begin{aligned}
ns(a_1)_{G_2\subset D_4}=ns(a_1)_{G_2\subset B_3}=&\{a_8,a_1,a_2\}\,,\\
ns(a_2)_{G_2\subset D_4}=ns(a_2)_{G_2\subset B_3}=&\{a_1,a_2,a_3,a_4,a_6,a_8\}\,,
\end{aligned}\quad + \quad a_{i}\to a_{i+2}\,,
\ee
where cyclic identification of the cluster variable indices is implied. Compared to the $G_2$ neighbour sets~\eqref{eq:observed_adjacencies} we notice that for the variables with odd indices $a_{2j-1}$ they remain the same, but for the variables with even indices $a_{2j}$ they double in size!

As a consequence, the embedding reduces the complement of the $G_2$ neighbour sets, the naive $\cN=4$ cluster adjacency predictions~\eqref{eq:N=4_G2_adjacency} from 40 to 20+8=28 relations. Specifically, it eliminates the following 12 of the restrictions of eq.~\eqref{eq:N=4_G2_adjacency},
\be
\cancel{\mathbf{A}'_{2i} \cdot \mathbf{A}'_{2j}{=}0}\,,\quad i\ne j=1,\ldots,4\,,
\ee
which the nonplanar triple ladder integral indeed does not obey. On the other hand, the embedding of $G_2$ inside the $B_3$ or $D_4$ cluster algebras still predicts that the restrictions of the second case in eq.~\eqref{eq:observed_adjacencies} should hold for the entire range of $i=1,\ldots 8$, whereas the integral obeys only half of these mixed odd/even index restrictions, for $i=3,\ldots 6$. In any case, we find it encouraging that the embedding procedure we have described modifies the naive $\cN=4$ cluster adjacency predictions~\eqref{eq:N=4_G2_adjacency} such that they approach the observed adjacency restrictions of the integral, and it would be interesting to explore if more general embeddings that go beyond what can be achieved by folding could also lead to a precise match.

In the rest of this subsection, we provide more details on how the result~\eqref{eq:B3D4inG2ENS} was obtained. This closely mirrors the $A_3\to C_2$ folding example provided in subsection~\eqref{subsec:FoldEmbed}.  A choice for the cluster variables and exchange matrix of the $D_4$ initial seed is
\begin{equation}
\{d_1,d_2,d_3,d_4\}\,,\quad    B_{D_4}=\begin{bmatrix}
        0 & -1 & 0 & 0\\
        1 & 0 & 1 & 1 \\
        0 & -1 & 0 & 0 \\
        0 & -1 & 0 & 0 
    \end{bmatrix}\,,
\end{equation}
which clearly is symmetric in the permutation of any of the rows 1,3,4. Following the folding procedure we described with respect to the first and fourth row, specifically eliminating the latter and the corresponding variable $d_4\to d_1$, yields the $B_3$ seed
\begin{equation}\label{eq:B3fromD4}
\{d_1,d_2,d_3\}\,,\quad    B_{B_3}=\begin{bmatrix}
        0 & -2 & 0\\
        1 & 0 & 1  \\
        0 & -1 & 0 
    \end{bmatrix}\,.
\end{equation}
Similarly, applying the same procedure for all rows 1,3,4, and in particular eliminating the latter two and their corresponding variables $d_4,d_3\to d_1$, leads to the $G_2$ seed
\begin{equation}\label{eq:G2fromD4}
\{d_1,d_2\}\,,\quad    B_{G_2}=\begin{bmatrix}
        0 & -3\\
        1 & 0   
    \end{bmatrix}\,.
\end{equation}
Indeed, the above exchange matrix is equivalent to the $G_2$ exchange matrix of eq.\eqref{eq:exchrank2} with $L=3$ up to a simultaneous reordering of rows and columns, which in turn amounts to an immaterial reordering of the variables of the cluster. In other words, one simply needs to relabel
\be\label{eq:G2relabelling}
d_1\to a_2\,, d_2\to a_1\,.
\ee
in order to match our previous conventions for the $G_2$ cluster algebra, as e.g. in eq.~\eqref{eq:G2alphabet} and figure~\ref{fig:exchG2}.

Notice that the $G_2$ seed~\eqref{eq:G2fromD4} may alternatively be obtained by folding the  $B_3$ seed~\eqref{eq:B3fromD4}, adding the third to the first column and eliminating $d_3\to d_1$. It is in this sense that $G_2$ is also contained in $B_3$ as depicted in figure~\ref{fig:exchB3}. Because the $B_3$ case is simpler, we will thus present the calculation of the embedded neighbour sets of $G_2$ with respect to the latter, only commenting on where the $D_4$ case differs.

The $B_3$ cluster algebra has 12 variables distributed to 20 clusters. Starting from the initial seed~\eqref{eq:B3fromD4}, and mutating according to the rules~\eqref{eq:matrixmutation}-\eqref{eq:variablemutation}, in our ordering conventions the cluster variables read,
\begin{align}
\Phi_{B_3}= & \left\{d_1,d_2,d_3,\frac{d_2+1}{d_1},\frac{d_3 d_1^2+1}{d_2},\frac{d_2+1}{d_3},\frac{d_3 d_1^2+d_2^2+2 d_2+1}{d_1^2 d_2},\frac{d_3 d_1^2+d_2+1}{d_2 d_3}, \right.\nonumber\\
& \left. \frac{d_3 d_1^2+d_2+1}{d_1 d_2},\frac{d_2^3+3 d_2^2+3 d_2+d_1^2 d_3+1}{d_1^2 d_2 d_3},\right.\\
& \left. \frac{d_3^2 d_1^4+3 d_2 d_3 d_1^2+2 d_3 d_1^2+d_2^3+3 d_2^2+3 d_2+1}{d_1^2 d_2^2 d_3},\frac{d_3 d_1^2+d_2^2+2 d_2+1}{d_1 d_2 d_3}\right\} \equiv \{d_{i}\}_{i=1}^{12}\,,\nonumber
\end{align}
whereas their neighbour sets are given by,
\be\label{eq:B3_NS}
\begin{aligned}
  & ns_{B_3}(d_1)= \left\{d_1,d_2,d_3,d_5,d_6,d_8\right\}\,, \\
  & ns_{B_3}(d_2)=\left\{d_1,d_2,d_3,d_4,d_6\right\}\,, \\
  & ns_{B_3}(d_3)= \left\{d_1,d_2,d_3,d_4,d_5,d_7,d_9\right\}\,,\\
  & ns_{B_3}(d_4)=\left\{d_4,d_2,d_3,d_6,d_7,d_{10}\right\} \,,\\
  & ns_{B_3}(d_5)=\left\{d_1,d_5,d_3,d_8,d_9\right\} \\
  & ns_{B_3}(d_6)=\left\{d_1,d_2,d_6,d_4,d_8,d_{10},d_{12}\right\} \,,\\
  & ns_{B_3}(d_7)= \left\{d_4,d_7,d_3,d_9,d_{10},d_{11},d_{12}\right\}\,,\\
  & ns_{B_3}(d_8)=\left\{d_1,d_5,d_8,d_6,d_9,d_{12},d_{11}\right\} \,,\\
  & ns_{B_3}(d_9)=\left\{d_5,d_9,d_3,d_7,d_8,d_{11}\right\} \,,\\
  & ns_{B_3}(d_{10})=\left\{d_4,d_{10},d_6,d_7,d_{12}\right\} \,,\\
  & ns_{B_3}(d_{11})=\left\{d_9,d_{11},d_7,d_8,d_{12}\right\} \,,\\  
  & ns_{B_3}(d_{12})=\left\{d_8,d_{12},d_6,d_{10},d_{11},d_7\right\}\,. \\  
  \end{aligned}
\ee
As mentioned already, the $G_2$ cluster algebra in our conventions is obtained from $B_3$ by the replacement $d_3\to d_1$, together with the relabeling~\eqref{eq:G2relabelling}. For the entire set of $B_3$ variables, this implies
\be\label{eq:B3varFolding}
\begin{aligned}
d_1,d_3&\to a_2\,,& d_2&\to a_1\,,& d_4,d_6&\to a_8\,,& d_5\to a_3\,, \\
d_7,d_{12}&\to a_6\,,& d_8,d_9&\to a_4\,,&d_{10}&\to a_7\,,&d_{11}\to a_{5}.
\end{aligned}
\ee

According to definition 1, in order to obtain the embedded neighbour set of a given $G_2$ variable, we need to take the union of the neighbour sets of all $B_3$ variables that reduce to it under the folding~\eqref{eq:B3varFolding}. For example, only $d_2$ reduces to $a_1$, and thus applying the replacement~\eqref{eq:B3varFolding} to its neighbour set in~\eqref{eq:B3_NS} gives,
\be
ns(a_1)_{G_2\subset B_3}=\{a_1,a_2,a_8\}\,.
\ee
On the other hand both $d_1$ and $d_3$ reduce to $a_2$, so in order to compute its embedded neighbour set one needs to perform the replacement~\eqref{eq:B3varFolding} on the union of the neighbour sets of the two $B_3$ variables, thus obtaining
\be
ns(a_2)_{G_2\subset B_3}=\{a_1,a_2,a_3,a_4,a_6,a_8\}\,.
\ee
Both of the last two formulas agree with what we already presented in~\eqref{eq:B3D4inG2ENS}, and the calculation proceeds also for the other neighbour sets in a similar fashion. Embedding $G_2$ inside $D_4$ also yields the same final result; the only difference in intermediate stages, is that the neighbour sets of $D_4$ variables that reduce to the same $G_2$ variables, also coincide.

\subsection{Adjacent $G_2$ polylogarithmic function counts}\label{subsec:function_counts}
Let us close this section by analysing the extent to which the adjacency restrictions~\eqref{eq:observed_adjacencies}, that we have discovered for the three-loop nonplanar ladder integral, reduce the size of the relevant function space. In other words, we will construct the space of polylogarithmic functions of (transcendental) weight $w$, satisfying the defining property,
\begin{equation}\label{eq:clusterfunctions}
    \dd f^{(w)}=\sum_{i}f_{i}^{(w-1)} \dd\log a_{i},
\end{equation}
where $f^{(w-1)}_{i}$ are weight-$(w-1)$ functions of the same type (with the recursion terminating with rational constants of weight $0$ on the right-hand side) and $a_i$ are the symbol letters, which in our case coincide the $G_2$ cluster variables~\eqref{eq:G2alphabet}. 

At weight $w$, this function space will contain the $\mathcal{O}(\epsilon^w)$ term in the expansion of all currently computed four-point one-mass integrals, when normalised such that this expansion starts at $\mathcal{O}(\epsilon^0)$. First constructing this space and then seeking to identify the integrals or even directly the physical quantities they contribute to, is at the heart of the perturbative analytic bootstrap programme for scattering amplitudes and beyond, see~\cite{Papathanasiou:2022lan} for a recent review. The success of this programme hinges on controlling the dimension of the function space at each weight, such that its construction is computationally feasible, and that the integral or physical quantity can be identified uniquely inside of it.

First of all, the dimension of the $G_2$ polylogarithmic function space at weight $w$ will not be $8^w$, because well-defined functions obey the property that double derivatives should yield the same result irrespective of the order of differentiation. This requirement is equivalent to the \emph{integrability condition},
\begin{equation}\label{eq:integrable}
\dd^2 f^{(w)}=0 \rightarrow \sum_i d f_i^{(w-1)} \wedge \dd\log a_i =0\,,
\end{equation}
which only allows particular weight-$(w-1)$ functions to appear on the right-hand side of eq.~\eqref{eq:clusterfunctions}. The construction of integrable polylogarithmic functions modulo transcendental constants, also known as symbols, has been automated in the \texttt{Mathematica} package \texttt{SymBuild}~\cite{Mitev:2018kie}, and applying it to our case yields the function counts shown on the first line of table~\ref{tab:intwords}.

\begin{table}
\centering
\begin{tabular}{llllllll}
\hline \hline
Weight & 1 & 2 & 3 & 4 & 5 & 6 & 7\\ \hline
No condition & 8 & 46 & 232 & 1093 & 4944 & 21790 & -\\ 
First entry & 3 & 14 & 61 & 262 & 1113 & 4700 & 19755\\ 
Adjacency & 3 & 14 & 54 & 196 & 684 & 2326 & 7796\\ \hline \hline
\end{tabular}
\caption{Dimension of the $G_2$ cluster function spaces (modulo transcendental constants) before and after constraints.}
\label{tab:intwords}
\end{table}

Then, a necessary condition such that the produced functions have physical branch cuts, as dictated by locality and unitarity, is the \emph{first entry condition}: The weight-one space must only contain letters that are Mandelstam variables. In the dimensionless alphabet of eq.~\eqref{eq:B1alphabet}, these correspond to the first three entries, and because of eq.~\eqref{eq:B1toG2_alphabet} this is also equivalent to,
\be
\text{First entry condition:}\,\, \vec f^{(1)}=\{\log z_1,\log z_2, \log (1-z_1-z_2)\}\sim \{\log a_2,\log a_4, \log a_8\}\,.
\ee
The function counts when the first entry condition is additionally imposed are shown in the second line of table~\ref{tab:intwords}\footnote{Note that the first entry condition is necessary but not sufficient condition for ensuring physical branch cuts. For the subspace of functions relevant for stress-tensor multiplet form factors in $\cN=4$ SYM theory, sufficient conditions were given in ~\cite{Dixon:2020bbt}, and were shown to further reduce the number of functions modulo transcendental constants. As our focus is to gauge the power of the adjacency restrictions~\eqref{eq:observed_adjacencies}, we will not consider such additional constraints here.}.

Finally, on top of integrability and the first entry condition, we may also impose the additional adjacency restrictions~\eqref{eq:observed_adjacencies}, with the corresponding function space dimensions shown in the third line of table~\ref{tab:intwords}. We notice that they start having an effect already at weight 3, and that by weight 6 they have reduced the size of the integrable $G_2$ functions obeying just the first entry condition by more than a half. These results provide strong indications that such adjacency restrictions may play an important role in future extensions of the bootstrap programme to four-point one mass integrals or Higgs plus jet amplitudes in the heavy-top limit.

\section{Conclusions and Outlook}\label{sec:OutlookConclusions}
We have demonstrated that all four-point one-mass master integrals through three loops computed to date are governed by a $G_2$ cluster algebra, enlarging the $C_2$ cluster algebra previously seen to be relevant at two loops. In particular, the alphabet~\eqref{eq:dimful_vars} entering their canonical differential equations~\eqref{eq:CDE_intro} was shown to be equivalent to the set of $G_2$ cluster variables~\eqref{eq:G2alphabet} thanks to the variable transformation~\eqref{eq:transform}. We find it remarkable that the $A_2,C_2$ and $G_2$ cluster algebras start becoming relevant at $L=1,2$ and 3 loops, respectively!

Focusing on the single integral with letters beyond those contained in the $C_2$ cluster algebra, shown in figure~\ref{fig:B1Topology}, we also looked for adjacency restrictions of the form $\mathbf{A}_{i} \cdot \mathbf{A}_{j}=0$ for the constant matrices entering the canonical differential equations. We discovered that using the $G_2$ cluster variable form of the alphabet reveals new adjacency restrictions, yielding a total of 20 instead of the 16 that were visible in the original alphabet. While the observed adjacency restrictions do not coincide with the naive $G_2$ cluster adjacency expectations seen to hold in $\mathcal{N}=4$ SYM theory, we showed that the two can be further aligned by embedding $G_2$ inside the larger $B_3$ or $D_4$ cluster algebras. We also illustrated the power of the adjacency restrictions we have observed by constructing the $G_2$ polylogarithmic function space, and noting that additionally imposing them leads to a significant reduction of its size.

Our work opens many exciting avenues for future inquiry. It would be very interesting to understand how the pattern of relevant cluster algebras continues for the complete set of master integrals at three as well as at higher loops, possibly entering the realm of infinite cluster algebras and the need to tame their infinite in a physically sensible manner, as was done in the case of cluster alphabets of $\cN=4$ SYM amplitudes~\cite{Arkani-Hamed:2019rds,Drummond:2019cxm,Henke:2019hve,Herderschee:2021dez,Henke:2021ity}. With respect to adjacency restrictions, in~\cite{Henn:2023vbd} it was pointed out that one of the tennis-court four-point one-mass master integrals, while still described by the $C_2$ alphabet, does not respect the subset of the observed adjacencies~\eqref{eq:observed_adjacencies} when restricted to this subalphabet. The authors of~\cite{He:2023umf} comment that Schubert analysis can be used to determine certain letters appearing in four-point one-mass master integrals, so it would be worthwhile to investigate if it could also provide any insight on adjacency restrictions. More importantly, could the cluster-algebraic structure of alphabets and adjacency restrictions be deduced from first principles, and employed to make new predictions? Recent progress on efficient methods for computing the Landau singularities of Feynman integrals~\cite{Panzer:2014caa,Fevola:2023kaw,Fevola:2023fzn,Helmer:2024wax}, and for also extracting symbol letters from them~\cite{Dlapa:2023cvx} or by related means~\cite{Heller:2019gkq,Heller:2021gun,Jiang:2024eaj,Matijasic:2024gkz}, makes us optimistic that this ambitious endeavour is within reach.

\acknowledgments
We'd like to thank William Torres Bobadilla for stimulating discussions. RA and GP acknowledge support from the Deutsche
Forschungsgemeinschaft (DFG) under Germany’s Excellence Strategy – EXC 2121 “Quantum Universe” – 390833306. GP's work was supported in part by the UK Science and Technology Facilities Council grant ST/Z001021/1 and by the Munich Institute for Astro-, Particle and BioPhysics (MIAPbP) which is funded by the DFG under Germany's Excellence Strategy – EXC-2094 – 390783311.

\bibliographystyle{JHEP}
\bibliography{bibliography.bib}
\end{document}